\documentstyle[11pt,aaspp4]{article}
\begin{document}
\title{Microlensing of an Elliptical Source by a Point Mass}
\author{David Heyrovsk\'y and Abraham Loeb } 
\medskip
\affil{Department of Astronomy, Harvard University, 60 Garden St.,
Cambridge, MA 02138}

\begin{abstract}

We present an efficient method for computing lightcurves of an elliptical
source which is microlensed by a point mass.  The amplification of an
extended source involves a two--dimensional integral over its surface
brightness distribution.  We show that for a general surface brightness
profile with an elliptical symmetry, this integral can be reduced to one
dimension.  We derive analytical results for the entire lightcurve in the
limit of low (e.g. planetary) mass lenses, and for the wings of all
microlensing lightcurves in general.  In both cases the lightcurve carries
information about deviations of the source from elliptical symmetry, e.g.
due to spots.  The method is used to find the amplification of a circular
red giant photosphere and an inclined accretion disk.  We demonstrate that
microlensing of an emission line from a disk can be used to infer the disk
velocity structure and surface brightness profile.

\end{abstract}


\centerline{To appear in {\it The Astrophysical Journal}, 1997}

\section{Introduction}

The sample of all known gravitational lenses is by now dominated by
stellar microlenses in our galaxy (Alcock et al. 1995, 1996; Udalski et al.
1994; Ansari et al. 1996; Alard 1996).  Even galactic {\it macro}-lenses of
quasars are often accompanied by microlensing events due to stars (Irwin et
al. 1989; Schild and Smith 1991; Schneider, Ehlers, \& Falco 1992 and
references therein), and thus emphasize that microlensing is a common
phenomenon.  Based on the fact that a solar mass lens at a distance $d$
offers an angular resolution $\sim (d/10 {\rm Gpc})^{-1/2}$micro-arcsec,
microlensing has long been recognized as a potential tool for resolving the
surface brightness distribution of compact sources.

Nevertheless, most stellar sources are smaller than the effective size of
their lenses. The finite extent of a stellar photosphere ($\sim R_\odot$)
is still much smaller than the Einstein radius of a solar mass lens [$\sim
10^3 R_\odot\times (M_{\rm lens}/M_\odot)^{1/2}(d/10{\rm kpc})^{1/2}$,
where $M_{\rm lens}$ is the lens mass], and so the lightcurve of galactic
microlenses typically carries little information about the brightness
distribution of the source (Paczy\'nski 1986; Udalski et al. 1994).  The
point source approximation breaks down only when the projected impact
parameter is comparable to the finite size of the source.  Although such 
circumstances are realized in a minority of all microlensing events, they
carry a large scientific payoff.  A spectroscopic identification of the
source size and distance in such a case, could break the parameter
degeneracy of the microlensing lightcurve and reveal the Einstein angle of
the lens (Loeb \& Sasselov 1995; Gould \& Welch 1996).  The appearance of
extended source signatures should be common when giant stars are
microlensed by planets (Loeb \& Sasselov 1995; Bennett
\& Rhie 1996).  By now, there are some examples of lightcurves with
unusually high amplification (i.e. small impact parameter) or short
duration (i.e. low lens masses) in which there are hints of finite
source signatures (Mao et al. 1994; Lennon et al. 1996; Bennett et al.
1996; Pratt 1996; Alcock et al. 1997).  
Searches for microlensing by planets (Albrow et al.
1996; Pratt et al.  1996) are likely to add more examples to this class of
events in the future. Spectroscopic monitoring of finite source events
could ultimately test the theory of stellar photospheres, especially for
giant stars (Loeb
\& Sasselov 1995; Sasselov 1996).

An analogous ordering of scales appears in stellar microlensing of the
continuum emission from quasars. For a star of mass $M_{\rm lens}$, the
Einstein radius at a cosmological distance is of order $\sim 5\times
10^{16}~(M_{\rm lens}/M_\odot)^{1/2}~{\rm cm}$.  In comparison, the optical
continuum emission in quasars is believed to originate from a compact
accretion disk. The UV bump observed in quasar spectra is often interpreted
as thermal emission at a disk temperature $T_{\rm disk}\sim 10^5 T_5~{\rm
K}$ (e.g. Laor 1990), and so the scale of the disk emission region must be
$\sim 10^{15}~T_5^{-2} L_{46}^{1/2}~{\rm cm}$, where $L_{46}$ is the
corresponding luminosity of the quasar in units of $10^{46}~{\rm
erg~s^{-1}}$.  Thus, for lens masses $M_{\rm lens}\gg 10^{-3}M_\odot$, the
continuum source is much smaller than the projected Einstein radius of the
lens.  This expectation is indeed confirmed in the lenses of Q2237+0305
(Irwin et al. 1989) and 0957+561 (Schild \& Smith 1991), where variability
due to microlensing has been detected (Wambsganss, Paczy\'nski, \&
Schneider 1990; Rauch \& Blandford 1991; Racine 1991; Gould \&
Miralda-Escud\'e 1996).

The above ordering of scales is reversed for the broad emission line
regions of quasars.  Reverberation studies of the time lag between
variations in the continuum and the line emission in active galactic nuclei
(AGN) indicate that the broad emission lines originate at a distance of
$\sim 3\times 10^{17}L_{46}^{1/2}~{\rm cm}$ (Peterson 1993; Maoz 1996).
This implies that the broad lines could be amplified through stellar
microlensing only for low--luminosity AGN ($L_{46}\la 10^{-2}$).
Microlensing of Seyfert galaxies offers a unique opportunity for resolving
the unknown structure and velocity distribution of the broad line region in
AGN.

For all of the above applications, it is important to find the
amplification in a variety of source and lens geometries. Evaluating the
net amplification of an extended source involves a two--dimensional
integral over its surface brightness distribution.  A direct calculation
of this integral raises some numerical problems (due to the divergence of
the amplification at the lens position) and is computationally time
consuming. Witt \& Mao (1994) expressed the amplification of a circular
source with uniform brightness in terms of elliptic integrals.
Alternatively, the amplification of uniform sources can be computed in the
image plane, avoiding the divergence. Realistic sources, however, have a
non--uniform surface brightness distribution (e.g.  due to limb--darkening
in stellar atmospheres) that also depends on the observed wavelength (Loeb
\& Sasselov 1995; Gould \& Welch 1996; Sasselov 1996). Witt \& Mao (1994)
proposed to sum over rings using the derivative of the brightness
distribution as the appropriate weight for their contributions to the
overall amplification. As a result, they traded one dimension in their
surface integration for the evaluation of their elliptic integral
solution. Other methods, such as image plane integration, require
approximating nonuniform sources as a superposition of uniform sources. In
this work, we propose another, more efficient approach to microlensing of
nonuniform sources. We show how to analytically reduce the surface
integral to one dimension for elliptically symmetric brightness profiles. 
This generalization can therefore be used to calculate microlensing of
inclined disks, in addition to its straightforward application to stellar
photospheres (Heyrovsk\'y, Loeb, \& Sasselov 1996). 

Microlensing is also a valuable tool for studying sources with organized
internal motions, such as rotating disks. If the gas in the disk emits a
spectral line, then the variation of the line profile during a
microlensing event would contain important information about the
trajectory of the lens and the inclination angle, velocity structure, and
brightness profile of the disk.  As the lens moves across the disk, it
amplifies the intensity at the Doppler shifted wavelength associated with
the rotation velocity behind the lens. By examining the time dependence of
the line profile it is possible, in principle, to deconvolve the radial
brightness profile of the disk down to the innermost radius sampled by the
lens trajectory.  Examples for relevant sources include the iron K$\alpha$
fluorescence line in AGN (Tanaka et al. 1995; Fabian et al. 1995; Iwasawa
et al. 1996), the maser lines emitted from compact disks in galactic
nuclei (e.g., Miyoshi et al. 1995), and the line emission from stellar
accretion disks (Warner 1995, and references therein).  A related effect
has previously been discussed in the context of the broad emission lines
in AGN (Nemiroff 1988; Wambsganss 1990; Schneider \& Wambsganss 1990),
although these lines are not believed to originate from a cold Keplerian
disk because they do not posses a double--peaked profile. 

In this work, we restrict our attention to microlensing by an isolated
point mass lens, and avoid more complex lensing environments (see, e.g.
Jaroszy\'nski, Wambsganss,
\& Paczy\'nski 1992; Wambsganss \& Kundic 1995).

The outline of the paper is as follows. In \S 2 we derive our main
mathematical results for the amplification of an elliptical source.  In \S
3, we derive analytical expressions for the entire lightcurve of a low mass
(e.g. planetary) lensing event (\S 3.1) and for the wings of the lightcurve
in general (\S 3.2).  Sections 4 and 5 are dedicated to specific
astrophysical sources. In \S 4 we calculate sample lightcurves for a giant
star with a realistic limb--darkened profile and for an inclined accretion
disk. In \S 5 we describe the time dependence of emission line profiles
from a Keplerian disk during a microlensing event.  Finally, \S 6
summarizes the main conclusions of this work.

\section{The Amplification Integral for Elliptical Sources}

In analysing the lensing geometry we use the two--dimensional sky
coordinates.  In this coordinate system, all projected distances are
angular.
 
We consider a general elliptical source with eccentricity $e$ and semimajor
axis $a$, which we set as our distance unit.  The location of the lens at a
given time can be defined by its displacement from the source center
$\tau_{0}$ and by the angle $\alpha_{0}$ between the source--lens
vector and the major axis of the source.  Figure 1 depicts the geometry
associated with our notations.  A lens with an Einstein radius 
$\epsilon$ amplifies the flux from a point source at a distance $\tau$
by a factor
\begin{equation}
A_{0}(\tau)=\frac{\tau^{2}+2\epsilon^{2}} 
{\tau\sqrt{\tau^{2}+4\epsilon^{2}}} \quad .
\label{eq:1}
\end{equation}
In the limit of geometric optics, an extended source can be viewed as a sum
of infinitesimal point--like sources. Its amplified flux can therefore be
obtained by weighting $A_{0}(\tau)$ with the surface brightness
distribution $B(\vec{r})$ and integrating over the projected surface area
of the source $\Sigma_{S}$,
\begin{equation}
F=a^{2}\int\limits_{\Sigma_S} B(\vec{r})\frac{\tau^2+2\epsilon^2}
{\tau\sqrt{\tau^2+4\epsilon^2}}\, d\Sigma\quad .
\label{eq:lensedflux}
\end{equation}
The net amplification is the ratio between this flux and
the flux of the source in the absence of the lens, 
\begin{equation}
A\equiv F/F_0\quad, 
\end{equation}
where
\begin{equation}
F_{0}=a^2\int\limits_{\Sigma_S} B(\vec{r})\, d\Sigma\quad .
\label{eq:fzero}
\end{equation}

It is most convenient to describe the surface brightness distribution of
the source in terms of a coordinate system which is centered on the
source. We set the $x$--axis along the major axis of the ellipse (see Fig.
1), and adopt the following coordinates: $x=\rho\cos\psi, y=\rho 
\sqrt{1-e^2} \sin\psi$. 
Curves of constant $\rho$ are nested ellipses with semimajor axes $\rho$
and equal eccentricity $e$ ; $\,\rho=1$ corresponds to the limb of the
source. Note that $\psi$, which runs from $0$ to $2\pi$, is {\it not} the
polar angle in the case of nonzero eccentricity. The area element is
$d\Sigma=\sqrt{1-e^2}\rho\,d\rho\,d\psi$, and so equation~(\ref{eq:fzero})
becomes
\begin{equation}
F_{0}=a^2\sqrt{1-e^2}\int\limits_0 \limits^{2\pi}\int\limits_0 
\limits^1  B(\rho,\psi)\rho\,d\rho\,d\psi\quad . 
\label{nolens}
\end{equation}

In calculating the lensed flux in equation~(\ref{eq:lensedflux}), it is
advantageous to use lens--centered polar coordinates for which the
amplification factor preserves its simple form~(\ref{eq:1}). If we erect
the corresponding $x_1$ axis from the lens towards the source center, we
obtain the coordinates: $x_1=\tau\cos\phi,\, y_1=\tau\sin\phi$ (see Fig.
1).  To express the source boundary and brightness distribution in terms of
$\tau$ and $\phi$, we need to transform the source--centered coordinates to
the lens--centered coordinates.  As seen from Figure 1,
\begin{equation}
\left( \begin{array}{c} x \\ y \end{array} \right) =
\left( \begin{array}{rc} -\cos\alpha_0 & \sin\alpha_0 \\ \sin\alpha_0 & 
\cos\alpha_0 \end{array} \right)
\left( \begin{array}{c} x_1-\tau_0 \\ y_1 \end{array} \right) ,
\end{equation}
or equivalently,
\begin{eqnarray}
\rho\cos\psi &=& -\tau\cos(\phi+\alpha_0)+\tau_0\cos\alpha_0 
\nonumber \\ \rho\sqrt{1-e^{2}}\sin\psi &=& 
\tau\sin(\phi+\alpha_0)-\tau_0\sin\alpha_0 \quad . 
\label{eq:6}
\end{eqnarray}
From these equations it is straightforward to express both $\rho$ and 
$\psi$ in terms of $\tau$ and $\phi$. For $\rho$ we get
\begin{equation}
\rho=\sqrt{\omega_1(\phi)\tau^2-\omega_2(\phi)\tau+\sigma_0^2}\quad ,
\label{eq:rho}
\end{equation}
with
\begin{eqnarray}
\omega_1(\phi)&=&1+\frac{e^2}{1-e^2}\sin^2(\phi+\alpha_0) \nonumber \\
\omega_2(\phi)&=&2\tau_0[\cos\phi+ 
\frac{e^2}{1-e^2}\sin\alpha_0\sin(\phi+\alpha_0)]  
\label{eq:8}\\
\sigma_0&=&\tau_0\sqrt{1+\frac{e^2}{1-e^2}\sin^2\alpha_0} \nonumber\quad .
\end{eqnarray}

The parameter $\sigma_0$ is the value of $\rho$ at the position of the lens
($\tau=0$). If the lens lies within the source boundary then $\sigma_0<1$,
and if it lies beyond the boundary, $\sigma_0>1$.  For a given angle $\phi$,
the integration limits in the $\tau$ direction are found by setting
$\rho=1$ in equation~(\ref{eq:rho}),
\begin{equation}
\tau_\pm(\phi)=\frac{\omega_2\pm\sqrt{\omega_2^2+4\omega_1(1-\sigma_0^2)}} 
{2\omega_1}\quad .
\label{eq:9}
\end{equation}
A line originating from the lens at an angle $\phi$ enters the source
boundary at $\tau_{-}(\phi)$ (provided $\sigma_0>1$) and exits at
$\tau_{+}(\phi)$.  If $\sigma_0<1$, $\phi$ runs from $-\pi$ to $\pi$. To
determine the $\phi$ integration limits in the $\sigma_0>1$ case, we set
$\tau_+(\phi_{\rm lim})=
\tau_-(\phi_{\rm lim})$ in equation~(\ref{eq:9}). After some algebra we find, 
\begin{eqnarray}
\label{eq:10}
\lefteqn{ \left( \begin{array}{c} \cos 2\phi_{\rm lim} 
\\ \sin 2\phi_{\rm lim} 
\end{array} 
\right) (\tau_0^4+e^4-2e^2\tau_0^2\cos 2\alpha_0) } \\ 
&=&(\tau_0^2+e^2-2) \left(
\begin{array}{c} \tau_0^2-e^2\cos 2\alpha_0 \\ e^2\sin 2\alpha_0 \end{array} 
\right) \pm 2\sqrt{\tau_0^2+e^2-e^2\tau_0^2\cos^2\alpha_0-1} \left(
\begin{array}{c} -e^2\sin 2\alpha_0 \\ \tau_0^2-e^2\cos 2\alpha_0 \end{array}
\right) . \nonumber
\end{eqnarray}
This expression generally yields four different solutions for $\phi_{\rm
lim}$, due to the implicit $\pm\pi$ ambiguity in this angle.  The two
values of interest are directed towards the source, and for them $\omega_2
(\phi_{\rm lim})\geq 0$. We use $\phi_-\in (-\pi,0)$ as the lower limit,
and $\phi_+\in (0,\pi)$ as the upper limit.  Since the area element in the
lens--centered coordinates is $d\Sigma=\tau\,d\tau\, d\phi$, the divergence
of the integrand in equation~(\ref{eq:lensedflux}) at $\tau=0$ is
eliminated. The explicit form of equation~(\ref{eq:lensedflux}) as a
function of the lens position is:
\begin{equation}
F(\tau_0,\alpha_0)=a^2\int\limits_{-\pi}\limits^{\pi} 
\int\limits_0\limits^{\tau_+(\phi)} B(\rho,\psi) 
\frac{\tau^2+2\epsilon^2} {\sqrt{\tau^2+4\epsilon^2}}~d\tau~d\phi
\quad , ~~~~~~~~~~~{\rm for}~~\sigma_0<1
\label{eq:sigmaless}
\end{equation} 
and
\begin{equation}
F(\tau_0,\alpha_0)=a^2\int\limits_{\phi_-}\limits^{\phi_+} 
\int\limits_{\tau_-(\phi)}\limits^{\tau_+(\phi)} B(\rho,\psi)
\frac{\tau^2+2\epsilon^2} {\sqrt{\tau^2+4\epsilon^2}}~d\tau~d\phi
\quad , ~~~~~~~~~~~{\rm for}~~\sigma_0>1. 
\label{eq:sigmamore}
\end{equation}

So far we made no assumptions about the brightness distribution
$B(\rho,\psi)$. In the following we concentrate on elliptical source
profiles with $B$ being independent of $\psi$. In particular, if the
brightness can be written as a series in powers of $\rho^2$
\begin{equation}
B(\rho)=\sum_{i=0}^{k}\beta_i\rho^{2i}\quad ,
\label{eq:bexpansion}
\end{equation}
then equation~(\ref{eq:rho}) 
can be used to transform this series to a polynomial in $\tau$,
\begin{equation}
B(\tau,\phi)=\sum_{L=0}^{k}\sum_{N=0}^{k-L}\frac{\omega_1^L(-\omega_2)^N} 
{L!\;N!}B^{(L+N)}\tau^{2L+N}\quad .
\label{eq:btauphi}
\end{equation}
Here the constants $B^{(n)}$ are the $n$--th derivatives of the brightness
profile~(\ref{eq:bexpansion}) with respect to $\rho^2$ at the position of 
the lens, 
\begin{equation}
B^{(n)}=\left[ \frac{d^n B}{d(\rho^2)^n}\right]_{\rho=\sigma_0} =
\sum_{i=n}^{k}\frac{i!}{(i-n)!}\beta_i\sigma_0^{2(i-n)}\quad .
\end{equation}
Note that in this expression the derivatives of the 
polynomial~(\ref{eq:bexpansion}) are formally evaluated at the lens 
position even if it lies beyond the source boundary. 

By substituting expression~(\ref{eq:btauphi}) into
equations~(\ref{eq:sigmaless}) and~(\ref{eq:sigmamore}), we 
are left with $\tau$--integrals of the general form
\begin{equation}
J_h(\mu,\nu)=\int\limits_{\mu}\limits^{\nu}\tau^h 
\frac{2\tau^2+4\epsilon^2} {\sqrt{\tau^2+4\epsilon^2}}\,d\tau\quad ,
\end{equation}
where $h=0,...,2k$. These integrals can be solved analytically. We get
\begin{equation}
J_0(\mu,\nu)=\left[\tau\sqrt{\tau^2+4\epsilon^2}\right]_{\mu}^{\nu} 
\quad ;
\label{eq:17}
\end{equation}
for $p\geq 0$
\begin{equation}
J_{2p+1}(\mu,\nu)=\frac{1}{2p+3} 
\left[\sqrt{\tau^2+4\epsilon^2}\left\{2\tau^{2p+2}-\frac{4\epsilon^2} 
{{2p \choose p}}\sum_{i=0}^p{2i \choose i}(-16\epsilon^2)^{p-i}\tau^{2i}  
\right\} \right]_{\mu}^{\nu} ;
\end{equation}
and for $p>0$ 
\begin{eqnarray}
J_{2p}(\mu,\nu)=&\frac{1}{p+1}
\left[\tau^{2p+1}\sqrt{\tau^2+4\epsilon^2}+4p{2p \choose p}  
(-\epsilon^2)^{p+1}\left\{{\rm arcsh} 
\frac{\tau}{2\epsilon} +\sqrt{\tau^2+4\epsilon^2}
\sum\limits_{i=1}\limits^p 
\frac{(-\epsilon^2)^{-i}}{2i {2i \choose i}} \tau^{2i-1}
\right\} \right]_{\mu}^{\nu}. \nonumber \\ 
& 
\label{eq:19}
\end{eqnarray}
In this way the two--dimensional integration over the source is 
reduced to a $\phi$--integral. 
Equations~~(\ref{eq:sigmaless}) and~(\ref{eq:sigmamore}) now read
\begin{equation}
F(\tau_0,\alpha_0)=\frac{a^2}{2}\int\limits_{-\pi}\limits^{\pi}
\sum_{L=0}^k\sum_{N=0}^{k-L}\frac{\omega_1^L(\phi)[-\omega_2(\phi)]^N} 
{L!\;N!} B^{(L+N)}J_{2L+N}(0,\tau_+(\phi))~d\phi ,
~~~~~~~~~~~{\rm for}~~\sigma_0<1, 
\label{eq:20}
\end{equation}
and 
\begin{equation}
F(\tau_0,\alpha_0)=\frac{a^2}{2}\int\limits_{\phi_-}\limits^{\phi_+}
\sum_{L=0}^k\sum_{N=0}^{k-L}\frac{\omega_1^L(\phi)[-\omega_2(\phi)]^N}
{L!\;N!} B^{(L+N)}J_{2L+N}(\tau_-(\phi),\tau_+(\phi))~d\phi , 
~~~~~{\rm for}~~\sigma_0>1 .
\label{eq:21}
\end{equation}
For completeness, we substitute equation~(\ref{eq:bexpansion})
into equation~(\ref{nolens}) to get the unlensed flux 
\begin{equation}
F_{0}=\pi a^2\sqrt{1-e^2}\sum_{i=0}^k\frac{\beta_i}{i+1} \quad .
\label{eq:23}
\end{equation}

Surface brightness profiles of astrophysical sources are often defined
through a set of discrete points $[\rho_i,B_i],\; i=1,\ldots ,n$, with
$\rho_1=0$ and
$\rho_n=1$. A straightforward application of the above results 
could be obtained through a least--squares fit of the
form~(\ref{eq:bexpansion}) to the data points.  However, a satisfactory
global fit for $\rho\in\langle0,1\rangle$ often requires the use of many
high--order terms in the series.  For the $k+1$ terms of
equation~(\ref{eq:bexpansion}) it is necessary to evaluate $2k+1$ functions
$J_0\cdots J_{2k}$, with increasing complexity at increasing $k$. As a 
result, an appropriate numerical integration scheme (such as Gaussian 
quadrature) would be more effective for high $k$. An alternative 
possibility is to interpolate between pairs of neighboring data points, 
so that for $\rho\in\langle\rho_i,\rho_{i+1}\rangle$ the brightness is 
described by an adequate function $B_i(\rho)$. A simple example for an 
interpolation scheme involves 
\begin{equation}
B_i(\rho)=a_i+b_i\rho^2\quad,
\label{eq:interpolation}
\end{equation}
where $a_i$ and $b_i$ are determined by the continuity conditions at
the boundary points,
\begin{equation}
B_i(\rho_i)=B_i,\quad B_i(\rho_{i+1})=B_{i+1},\quad i=1\ldots n-1\;.
\label{eq:continuity}
\end{equation}
This interpolation scheme for the brightness profile does not have a
continuous derivative at the data points $\rho_i$. Smoothness can be easily 
achieved by adding a $\rho^4$ term to equation~(\ref{eq:interpolation}) and
by adding conditions for the continuity of the derivative in 
equation~(\ref{eq:continuity}).

Following the interpolation scheme of equation~(\ref{eq:interpolation}),
the source is divided into $n-1$ concentric bands, bounded by ellipses with
semimajor axes $\rho_i$. The brightness profile inside the $j$--th band
is described by values $a_j$ and $b_j$. A line of constant angle
$\phi$ from the lens passes through a sequence of $l$ bands in the
$\tau$--direction, and crosses their boundaries at values
$\tau_1<\tau_2<\cdots<\tau_{l+1}$ (see Fig. 2).  We correspondingly
substitute in the integrand of equation~(\ref{eq:20})
\begin{equation}
B^{(L+N)}J_{2L+N}(0,\tau_+) \longrightarrow\sum_{i=1}^l B_{s_i}^{(L+N)} 
J_{2L+N}(\tau_i,\tau_{i+1})\quad,
\end{equation}
where $\tau_1=0,\;\tau_{l+1}=\tau_+$, and $s_i$ is the index of the band
crossed between $\tau_i$ and $\tau_{i+1}$. An analogous substitution with
$\tau_1=\tau_-$ should be performed in the integrand of
equation~(\ref{eq:21}). 

The sequence of bands that are crossed can easily be determined from the
geometry illustrated in Figure 2.  If the lens lies outside the source
boundary ($\sigma_0>1$), one needs to find the innermost band which is
crossed.  The $\rho$--interval of this band contains the minimum value of
$\rho$ which is reached along the $\phi=const$~ ray. Equation~(\ref{eq:rho})
yields
\begin{equation}
\rho_{min}=\sqrt{\sigma_0^2-\frac{\omega_2^2}{4\omega_1}}\; .
\end{equation}
If the lens lies inside the source boundary ($\sigma_0<1$), the sequence of
bands starts at the lens position ($\rho=\sigma_0$), and the band with
$\rho_{min}$ has to be determined only if $\omega_2>0$. To calculate the
crossing points $\tau_i$, it is then sufficient to know that intersections
with ellipse $\rho_j$ occur at
\begin{equation}
\tau=\frac{\omega_2\pm\sqrt{\omega_2^2+4\omega_1(\rho_j^2-\sigma_0^2)}}
{2\omega_1}\quad .  
\end{equation}
If the necessary number of bands is high, numerical evaluation of the 
$\tau$ integral should be considered as an alternative.

The method has so far been presented for an elliptical source.  Most of the
equations are considerably simpler for a circular source. By setting the
eccentricity to zero, $e=0$, the source--centered coordinates $\rho,\psi$
become regular polar coordinates, with $a$ being the source radius.
Equations~(\ref{eq:8}) reduce to: $\omega_1=1$,
$\omega_2(\phi)=2\tau_0\cos{\phi}$, and $\sigma_0=\tau_0$.  The $\tau$
integration limits in equation~(\ref{eq:9}) are
\begin{equation}
\tau_\pm=\tau_0\cos{\phi}\pm\sqrt{1-\tau_0^2\sin^2{\phi}}\quad ,
\end{equation}
and the angular integration limits in equation~(\ref{eq:10}) are given by 
\begin{equation} 
\phi_\pm=\pm\arcsin{\frac{1}{\tau_0}}\quad .
\end{equation}
Due to the circular symmetry, the angular integration in
equations~(\ref{eq:20}) and~(\ref{eq:21}) can be performed over half
the interval and the result multiplied by a factor of two.

In principle, our method can also be used to describe the effect of spots
on lightcurves of extended sources (Sasselov 1996). Since the observed flux
is linearly dependent on the source brightness, one could simply
superimpose the lightcurve of an elliptical spot on top of the background
lightcurve.

\section{Analytical Results}
\subsection{Low Mass Lenses}

The results obtained for brightness profile~(\ref{eq:bexpansion}) can be
used to study the limiting case when the Einstein radius of the lens is
much smaller than the source size ($\epsilon\ll 1$). The lensed flux in
equations~(\ref{eq:20}) and~(\ref{eq:21}) depends on $\epsilon$ only
through the functions $J_h$ in equations~(\ref{eq:17})--(\ref{eq:19}).  For
small $\epsilon$, these functions can be expanded as power series in
$\epsilon$. We make use of the expansions
\begin{eqnarray}
\sqrt{\tau^2+4\epsilon^2}&=&\tau+\frac{2\epsilon^2}{\tau}-
\frac{2\epsilon^4}{\tau^3}+O(\epsilon^6) \nonumber \\
{\rm arcsh}\frac{\tau}{2\epsilon}&=&-\ln\epsilon+\ln\tau+  
\frac{\epsilon^2}{\tau^2}+O(\epsilon^4)\quad .
\end{eqnarray}
If the lens position is inside the source boundary ($\sigma_0<1$), we obtain
\begin{eqnarray}
&J_0(0,\tau_+)&=\tau_+^2+2\epsilon^2 
-\frac{2\epsilon^4}{\tau_+^2}+o(\epsilon^4) \nonumber \\
&J_1(0,\tau_+)&=\frac{2}{3}\tau_+^3+\frac{8\epsilon^3}{3} 
-\frac{4\epsilon^4}{\tau_+}+o(\epsilon^4)  \\
&J_2(0,\tau_+)&=\frac{\tau_+^4}{2}-4\epsilon^4\ln\epsilon
+(4\ln\tau_+ -3)\epsilon^4+o(\epsilon^4) \nonumber \\
L>2\quad&J_L(0,\tau_+)&=\frac{2}{L+2}\tau_+^{L+2}+\frac{4\epsilon^4}{L-2}
\tau_+^{L-2}+o(\epsilon^4) \nonumber\quad .
\end{eqnarray}
If the lens is outside the source ($\sigma_0>1$), we get
\begin{eqnarray}
L\neq2\quad&J_L(\tau_-,\tau_+)&=\left[\frac{2}{L+2}\tau^{L+2} 
+\frac{4\epsilon^4}{L-2}\tau^{L-2}\right]_{\tau_-}^{\tau_+} 
+o(\epsilon^4) \nonumber \\
&J_2(\tau_-,\tau_+)&=\left[\frac{\tau^4}{2} 
+4\epsilon^4\ln\tau\right]_{\tau_-}^{\tau_+}+o(\epsilon^4) 
\quad .
\end{eqnarray}
The observed flux can now be obtained from equations~(\ref{eq:20})
and~(\ref{eq:21}):
\begin{equation}
F(\tau_0,\alpha_0)=F_0+2\pi a^2\epsilon^2B^{(0)} 
+O(\epsilon^4\ln\epsilon)~~~~~~~~~~~~~~~{\rm for}~~\sigma_0<1,
\label{eq:35}
\end{equation}
and
\begin{equation}
F(\tau_0,\alpha_0)=F_0+2\epsilon^4 a^2\int\limits_\Sigma 
\frac{B}{\tau^4}\,d\Sigma+O(\epsilon^6)\quad~~~~~~~~~~~~~~{\rm for}~~
\sigma_0>1 .
\label{eq:36}
\end{equation}

Since $B^{(0)}$ is the brightness at the position of the lens, the
lightcurve traces the brightness profile along the lens trajectory to
leading order ($\epsilon^2$). For $\sigma_0<1$, the excess flux equals
twice the Einstein ring area times $B^{(0)}$.  The accuracy of this
leading--order approximation will be demonstrated through a numerical
example in \S 4. We expect the first two terms in equations~(\ref{eq:35})
and~(\ref{eq:36}) to hold in general for an arbitrary surface brightness
distribution, as long as the variations in $B({\vec{r}})$ are weak on the
scale of the Einstein radius of the lens.
Note that the convergence of the above expansions requires that
$2\epsilon$ be smaller than $\tau_-$ and $\tau_+$. Hence, the derived
expression for the lightcurve is valid as long as the lens lies a distance
greater than $2\epsilon$ away from the source boundary.

\subsection{Lightcurve Wings}

We next study the effect that a finite source size has on the wings of a
general lightcurve. For this purpose we start from
equation~(\ref{eq:lensedflux}) and place no restrictions on the brightness
distribution $B(\vec{r})$.  We associate the ``wings'' of the lightcurve
with the region over which the lens--source separation is larger than both
the Einstein diameter of the lens and the size of the source, i.e. we
assume that $\tau_0\gg 2\epsilon$ and $\tau_0\gg 1$.  Rewriting
equation~(\ref{eq:lensedflux}) in terms of source coordinates, we get
\begin{equation}
F=a^2\sqrt{1-e^2}\int\limits_0 \limits^{2\pi}\int\limits_0
\limits^1 B(\rho,\psi)\left(1+\frac{2\epsilon^2}{\tau^2}\right) 
\left(1+\frac{4\epsilon^2}{\tau^2}\right)^{-\frac{1}{2}} 
\rho\,d\rho\,d\psi \quad ,
\label{eq:37}
\end{equation} 
where $\tau^2$ can be expressed from equations~(\ref{eq:6}),
\begin{equation}
\tau^2=\tau_0^2-2\rho\tau_0(\cos\alpha_0\cos\psi- 
\sqrt{1-e^2}\sin\alpha_0\sin\psi)+\rho^2(1-e^2\sin^2\psi) \quad .
\label{eq:38}
\end{equation}
The amplification factor in the integrand of equation~(\ref{eq:37}) can 
be expanded in powers of $\tau_0^{-1}$, 
\begin{equation}
\left(1+\frac{2\epsilon^2}{\tau^2}\right) 
\left(1+\frac{4\epsilon^2}{\tau^2}\right)^{-\frac{1}{2}} 
=1+\frac{2\epsilon^4}{\tau_0^4}+\frac{8\rho\epsilon^4}{\tau_0^5} 
(\cos\alpha_0\cos\psi-\sqrt{1-e^2}\sin\alpha_0\sin\psi)+O(\tau_0^{-6})\; .
\end{equation}
By substituting this result into equation~(\ref{eq:37}) we obtain,
\begin{equation}
F=\left(1+\frac{2\epsilon^4}{\tau_0^4}\right) F_0+ 
\frac{8\epsilon^4a^2\sqrt{1-e^2}}{\tau_0^5}\int\limits_0\limits^1 
\int\limits_0\limits^{2\pi} B\, 
(\cos\alpha_0\cos\psi-\sqrt{1-e^2}\sin\alpha_0\sin\psi)\rho^2\,d\psi\, 
d\rho + O(\tau_0^{-6}) .
\label{eq:40}
\end{equation}

The leading correction to the unlensed flux (of order $\tau_0^{-4}$) is
independent of the source structure, and appears also for a point source.
The next term (of order $\tau_0^{-5}$) vanishes for a point source, as
well as for any symmetric source with $B(\rho,\psi)=B(\rho)$.  The
integral in this term is, in fact, the projection of the ``brightness
dipole moment'' $\int\vec{r}B(\vec{r})\,d\Sigma$ on the source--lens
direction. This term therefore reflects the asymmetry in the brightness
distribution of the source, which could result from the existence of hot
or cold spots on its surface (Sasselov 1996; Gould \& Miralda-Escud\'e
1996). The term vanishes if we redefine the positions relative to the 
center of brightness, rather than the geometrical center of the 
source. 

\section{Sample Lightcurves of Astrophysical Sources}

The microlensing lightcurve of the source is obtained from the time
dependence of the lens position coordinates, $\tau_0$ and $\alpha_0$. A
linear motion of the lens with respect to the source can be described by
two parameters: the angle $\beta$ between the major axis of the source and
the lens trajectory, and the impact parameter $p$ (see Fig. 3). To avoid
ambiguity, we assign a negative sign to $p$ if the source center lies to the
right of the lens trajectory. For convenience, we normalize the time $t$ in
units of semimajor axis crossing--time and set $t=0$ at closest approach.
The geometry illustrated in Figure 3 then yields
\begin{eqnarray}
\tau_0&=&\sqrt{p^2+t^2} \\
\left( \begin{array}{c} \cos\alpha_0 \\ \sin\alpha_0 \end{array} \right) 
&=& \frac{1}{\tau_0} \left( \begin{array}{cc} \cos\beta & \sin\beta \\ 
-\sin\beta & \cos\beta \end{array} \right) 
\left( \begin{array}{c} t \\ p \end{array} \right) \quad .
\end{eqnarray}
In the case of a circular source (e.g. a lensed star), only $p$ needs to
be specified and the angles $\beta$ and $\alpha_0$ are redundant.

As an example of stellar microlensing we present a theoretical lightcurve
for the MACHO 95--30 event (Pratt 1996, Alcock et al. 1997). The source, a
red giant in the galactic bulge, was amplified by a lens with an Einstein
radius $\epsilon\approx 13$ and an impact parameter $p\approx 0.7$.  The
R--band brightness profile shown in Figure 4 is based on an atmosphere
model calculation of the source by Sasselov (1996). The computed
lightcurve is plotted in Figure 5 together with a point-source lightcurve,
for comparison. 

Figure 6 compares the exact lightcurve of a low mass lens with the leading
order term in the expansion described by equation~(\ref{eq:35}).
In difference from the real MACHO 95--30 event, we assume in this
example that the lens has an Einstein radius $\epsilon=0.1$ and an
impact parameter $p=0.2$. In this case, the
leading order term (of order $\epsilon^2$) in equation~(\ref{eq:35})
captures the main characteristics of the lightcurve to a reasonable
accuracy.

The case of an elliptical source can be illustrated through the example of
microlensing of an AGN disk. A circular disk which is inclined by an angle
$i$ relative to the line-of-sight, would appear as an ellipse with
eccentricity $e=\sin{i}$ on the sky projection.  The semimajor axis would
then equal the disk radius, and the $\rho = const$ contours would be
elliptical projections of circles, with $\psi$ being the standard polar
angle on the face of the disk. In calculating a sample lightcurve for this
application, we use the brightness profile of a stationary thin accretion
disk (see Frank, King, \& Raine 1985) with
$B(\rho)\propto\left[1-\sqrt{{\rho_{in}}/{\rho}}\right]\rho^{-3}$.  We
assume a disk with an inner radius $\rho_{in}=0.1$ and an inclination
$i=60^\circ$, which is lensed by a lens with an Einstein radius
$\epsilon=5$.  Figure 7 contains lightcurves for events with the same
impact parameter $p=0.8$ but different approach angles $\beta$.  While the
lightcurves of a circularly symmetric source are invariant to $\beta$, the
inclination of the disk leaves a distinct $\beta$--dependent signature on
the amplitude and symmetry of its lightcurve.

\section{Microlensing of an Emission Line From a Keplerian Disk}

Next we consider the effect of microlensing on the spectral shape of an
emission line from an accretion disk. Because of the Doppler shift
associated with the disk rotation, the amplification as a function of
wavelength could help unravel the inclination and surface brightness
distribution of the disk.  

To examine the microlensing distortion of the line profile we adopt the
simplest model of a planar thin Keplerian disk with an inner radius
$\rho_{in}$ and inclination $i$. We parametrize the disk surface by the
coordinates $\rho,\,\psi$ which were introduced in \S 2, and assume a
Keplerian velocity profile $v(\rho)=v_{in}\sqrt{{\rho_{in}}/{\rho}}$, where
$v_{in}$ is the velocity at the inner edge of the disk.  We assume that the
cold gas in the disk emits a line at a rest--frame wavelength $\lambda_0$,
and with a thermal width that is much smaller than the Doppler width
induced by the rotation of the disk.  We therefore approximate the
rest--frame line profile by a delta--function centered on $\lambda_0$. For
simplicity, we neglect any relativistic effects on the radiation emitted by
the disk.  Due to variations of the line-of-sight velocity across the
source, the Doppler-shifted wavelength reaching the observer from a point
$(\rho,\,\psi)$ is
\begin{equation}
\lambda=\lambda_0\left[ 1+\frac{v(\rho)}{c}\sin{i}\cos{\psi}\right]\quad ,
\end{equation}
where $c$ is the speed of light.  In this notation, $\psi=0$ is on the
redshifted half of the disk.  The spectral intensity profile of the line
emitted by a microlensed disk is then
\begin{equation}
I(\lambda)=\int\limits_{\rho_{in}}\limits^1\int\limits_{-\pi}\limits^\pi 
j(\rho)\,\delta\left[\lambda-\lambda_0\left(1+\frac{v(\rho)}{c}\sin{i} 
\cos{\psi}\right)\right] A_0(\tau(\rho,\psi))\,\rho\cos{i}\,d\psi\,d\rho 
\quad ,
\label{eq:42}
\end{equation}
where $j(\rho)$ is the spectral emissivity and $A_0(\tau)$ is the
amplification factor in equation~(\ref{eq:1}). The displacement 
$\tau(\rho,\psi)$ of the lens from the integration point is given by
equation~(\ref{eq:38}) with $e=\sin{i}$.  We normalize the wavelength shift
by its maximum value, ${(v_{in}/c)\sin{i}}$, and denote the normalized
fractional shift by
$\Delta=\left({\lambda}/{\lambda_0}-1\right)/[{(v_{in}/c)\sin{i}}]$.
Integration of the delta--function in equation~(\ref{eq:42}) over $\psi$
yields
\begin{equation}
I\left(\lambda_0[1+\Delta\frac{v_{in}}{c}\sin{i}]\right)=\int 
\limits_{\rho_{in}}\limits^{\rho_{max}}j(\rho)\, 
\frac{A_0\left(\tau(\rho,\arccos[\Delta\sqrt{\rho/\rho_{in}}])\right) 
+A_0\left(\tau(\rho,-\arccos[\Delta\sqrt{\rho/\rho_{in}}])\right)} 
{\lambda_0 (v_{in}/c)\,\sqrt{{\rho_{in}}/{\rho}-\Delta^2}~\tan{i}}\,\rho 
\,d\rho . 
\label{eq:43}
\end{equation}
The two terms in the numerator reflect the fact that points with angles
$\psi$ and $-\psi$ have the same line-of-sight velocity. The upper
integration limit depends on wavelength. For small wavelength shifts
$|\Delta|\le\sqrt{\rho_{in}}$, one gets contributions from all radii so
that $\rho_{max}=1$. Larger shifts $|\Delta|\ge\sqrt{\rho_{in}}$ originate
only from the inner disk, with $\rho_{max}=\rho_{in}/\Delta^2$.

For concreteness, we assume a power--law profile for the disk emissivity,
$j(\rho)\propto\rho^{n}$. Figure 8 shows the sensitivity of the {\it
unlensed} line profile to the power--law index $n$. For the purpose of
this illustration, we have kept the total luminosity of the disks
constant, namely $\int j(\rho)\rho\,d\rho=const$. The unlensed profiles
show two characteristic peaks at $|\Delta|=\sqrt{\rho_{in}}$. We have
chosen $\rho_{in}=0.01$ and $i=60^\circ$ in the examples shown. As is
apparent, the more centrally concentrated the emissivity law is, the more
pronounced the line wings are and the less pronounced the peaks are.  This
tendency results from the enhanced weight that the inner region of the
disk obtains in the intensity integral~(\ref{eq:43}), as $n$ gets more
negative.  The wings of the line correspond to high rotation speeds which
are generically obtained in the inner region of the disk. 

Figure 9 shows the effect of lensing on the line profile for $n=-1$.  In
this calculation we assumed a lens with an Einstein radius
$\epsilon=1$, located at several different positions along the major
axis of the source ($\alpha_0=0$). For clarity, we show results only for
the redshifted side of the disk. Apart from an overall amplification of the
total flux in the line, lensing introduces an asymmetry between the
intensities of the two peaks. The position of the peaks remains unaffected.
The curve with $\tau_0=0.1$ illustrates another feature, present when the
lens is positioned on top of the disk. In this situation, the profile
diverges at the emission wavelength of the geometric point behind the lens,
$\Delta=\sqrt{\rho_{in}/\rho_{\rm lens}}\cos{\psi_{\rm lens}}$.  This
divergence is integrable, and so a filter with a finite wavelength
resolution would register it as a third peak. In reality, the formal
divergence will be smoothed out by the thermal width of the line. The
time--dependent intensity profile around the moving third peak can be used
to map the axisymmetric brightness profile of the disk as the lens scans
the source along its linear trajectory.

The asymmetry of the microlensed line profiles in Figure 9 indicates a 
change in the mean redshift of the emitted line, depending on the lens 
position. This effect has been previously studied for various models of 
the broad line regions of quasars (Nemiroff 1988; Schneider \& Wambsganss 
1990). The redshift change $\Delta z$ can be simply computed from the spectral 
intensity $I$ in equation~(\ref{eq:43}) as 
\begin{equation}
\Delta z =\frac{\int\limits_{-1}^{1}I\Delta\,d\Delta} 
{\int\limits_{-1}^{1}I\,d\Delta}\; \frac{v_{in}}{c}\, \sin{i}\, .
\end{equation}
Figure 10 shows the change in redshift as a function of lens position
along the major axis of the disk, for the same situation as in Figure 9. 
In this case the redshift can change by more then 10\% of the maximum 
Doppler shift from the inner edge of the disk, if the lens is positioned 
close to the inner edge.

\section{Discussion}

We have developed an efficient method for computing microlensing
lightcurves of sources with an elliptical symmetry and a general surface
brightness distribution. Equations~(\ref{eq:17})--(\ref{eq:23}) express
the amplification of elliptical sources in terms of a one--dimensional
integral. This integral is considerably simpler for a circular source. 
The method is well suited for the study of spectral changes due to
microlensing. Such changes have already been observed in the MACHO 95--30
event (Alcock et al. 1997).  We are currently using the technique for
modelling these observations (Sasselov, Heyrovsk\'y \& Loeb 1997). The
method can also be easily extended to describe the effect of elliptical
spots on the background lightcurve.We have derived fully analytical
results in the limit of low lens masses [cf. equations~(\ref{eq:35}),
(\ref{eq:36})] and for the wings of the lightcurve in general
[equation~(\ref{eq:40})]. In both cases, the lightcurve carries
information about deviations of the source from elliptical symmetry (e.g.,
due to hot or cold spots).  The low mass limit case applies to lensing of
giant stars in the Milky Way bulge by planets of mass $M_{\rm pl}$, for
which the Einstein radius is $\sim 3 R_\odot\times (M_{\rm pl}/
M_\oplus)^{1/2}$. The lightcurve in this case obtains the form of a low
amplitude increase in the source intensity over the source crossing time.
The amplitude of the excess flux is twice the area of the Einstein ring
times the brightness of the source at the lens position [cf.
equation~(\ref{eq:35})].  Events of this type might soon be discovered
through ongoing microlensing searches for planets (Albrow et al.  1996;
Pratt et al. 1996). Finite source signatures appear whenever the impact
parameter is comparable to the source size, including high amplification
events caused by massive lenses.  Figure 5 shows the theoretical
lightcurve corresponding to the source and lens parameters of a
microlensing event of this type, MACHO 95--30 (Pratt 1996, Alcock et al.
1997).  The detection rate of events with finite source signatures is
expected to increase if future searches will focus on the source
population of giant stars (Gould 1995a).  When such signatures are
combined with spectroscopic identification of the sources, they can be
used to break the degeneracy of microlensing lightcurves (Loeb \& Sasselov
1995; Gould \& Welch 1996). 

Our formalism is particularly suitable for the analysis of microlensing of
inclined accretion disks which appear elliptical in projection on the sky.
The microlensing lightcurve of the disk depends on its surface brightness
distribution and inclination (cf. Fig. 7). If a thin disk emits a spectral
line, then it is possible to deconvolve its properties from the variation
of the line profile during a microlensing event. When the lens position
overlaps with the disk, a third peak is added to the standard
``double--peak'' profile of the disk emission. The microlensing peak
appears at the emission wavelength corresponding to the Doppler velocity
behind the lens (cf. Fig. 9). As the lens moves along a straight line, the
temporal evolution of the intensity and wavelength of the third peak can be
used to map the disk structure.  Thus, microlensing of a spectral line
offers a unique opportunity for unraveling the surface brightness and
velocity profiles of accretion disks.  The main practical obstacle for such
mappings is the potential for intrinsic variability in the disk properties
during the event. Figure 10 illustrates the change in the mean redshift of 
the spectral line, due to the asymmetry of the microlensed profiles. The 
change can exceed 10\% of the maximum Doppler shift in the disk. 

The signatures described in this paper should be more complicated in
environments where the microlensing optical depth is not small, such as the
cores of galactic {\it macro}-lenses at cosmological distances.  However,
our {\it isolated point lens} calculation still applies to the outer halos of
such galaxies, where the optical depth is low. Quasar absorption spectra
can be used to select such foreground galactic halos. In particular, damped
Ly$\alpha$ or metal--line absorption features are often associated with
projected impact parameters of $\sim 10$--$100$ kpc from galactic centers
(Steidel et al. 1994, 1995, 1996), and could therefore be used to signal an
intervening galactic halo. If galactic halos are composed of Massive
Compact Halo Objects, then a significant fraction of all damped Ly$\alpha$
absorbers should show evidence for isolated microlensing events (Perna \&
Loeb 1997), of the type discussed in this work. Isolated microlensing
events from intergalactic stars are also possible but only for a minority
of all quasars (Dalcanton et al. 1994; Gould 1995b).

\acknowledgements
We thank Dimitar Sasselov for providing the brightness profile of the
MACHO 95--30 source, and Robert Nemiroff for suggesting the calculation of
the change in the mean redshift of microlensed emission lines. We also thank 
both for valuable comments on the manuscript. This work was supported in part
by the NASA ATP grant NAG5-3085 and the Harvard Milton fund (for AL).

\begin{figure}
\plotone{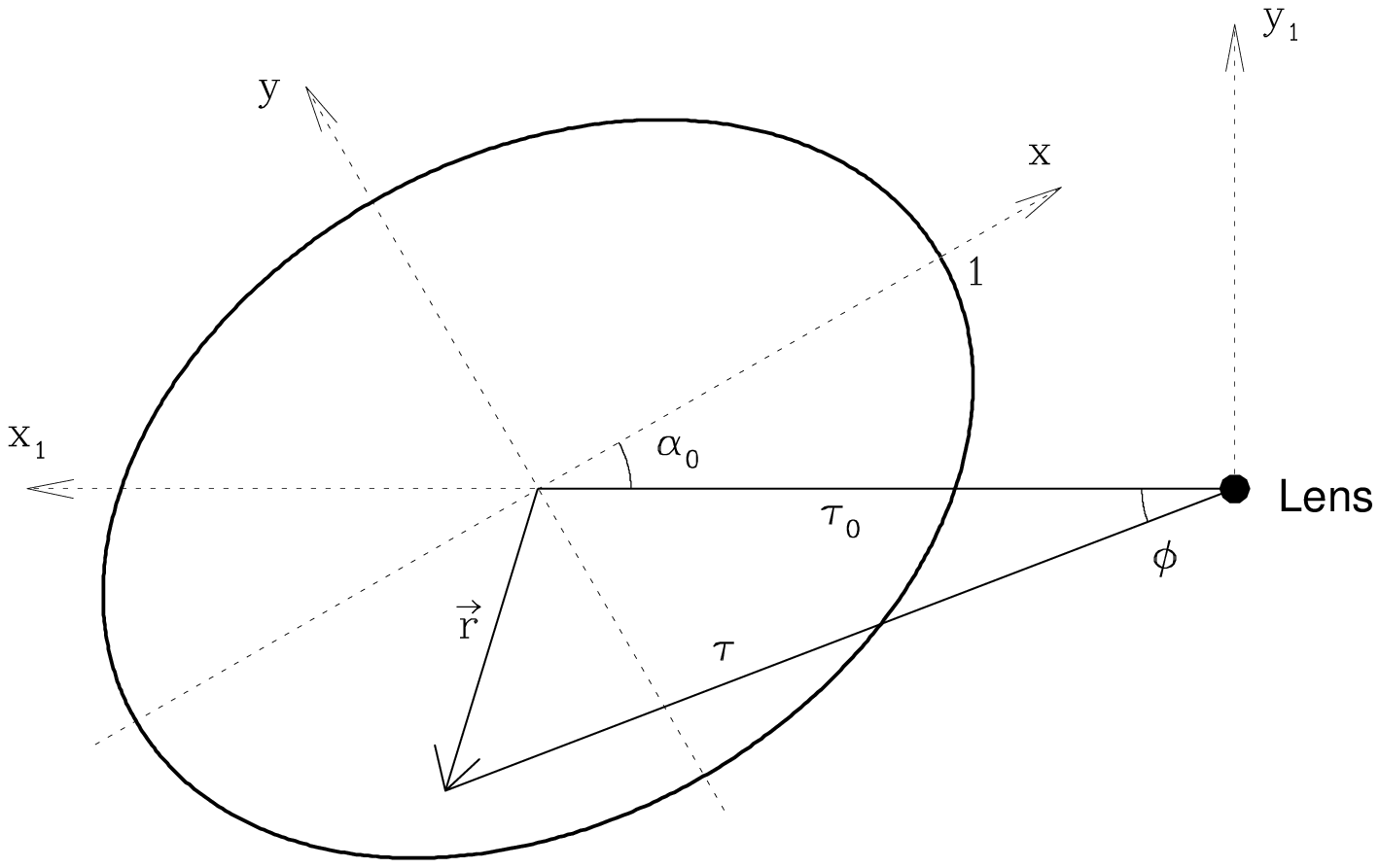}
\caption{Geometry of the lens and an elliptical source projected on 
the sky.  $\vec{r}$ marks an arbitrary point on the source.}
\end{figure}

\begin{figure}
\plotone{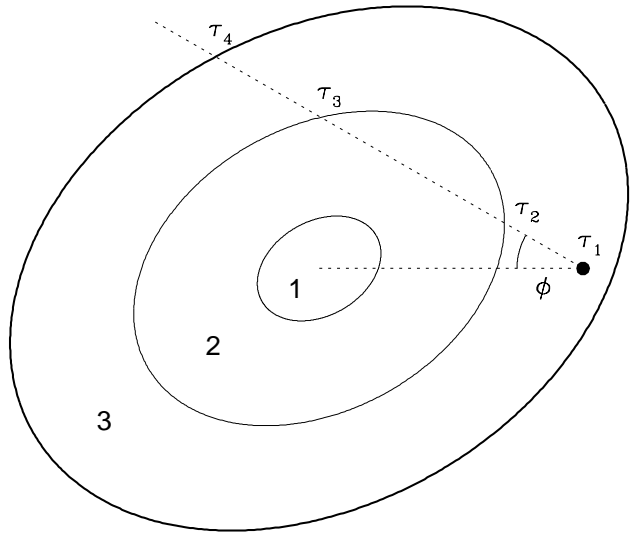}
\caption{Sketch of the integration method geometry. The source is 
divided into three bands, each of which is characterized by its separate
section of the brightness profile interpolation. For the angle $\phi$ in
this figure, the band sequence is $\{s_1, s_2, s_3\}=\{3, 2, 3\}$ and
$\tau_1=0\, ,\,\tau_4=\tau_+(\phi)$.}
\end{figure}

\begin{figure}
\plotone{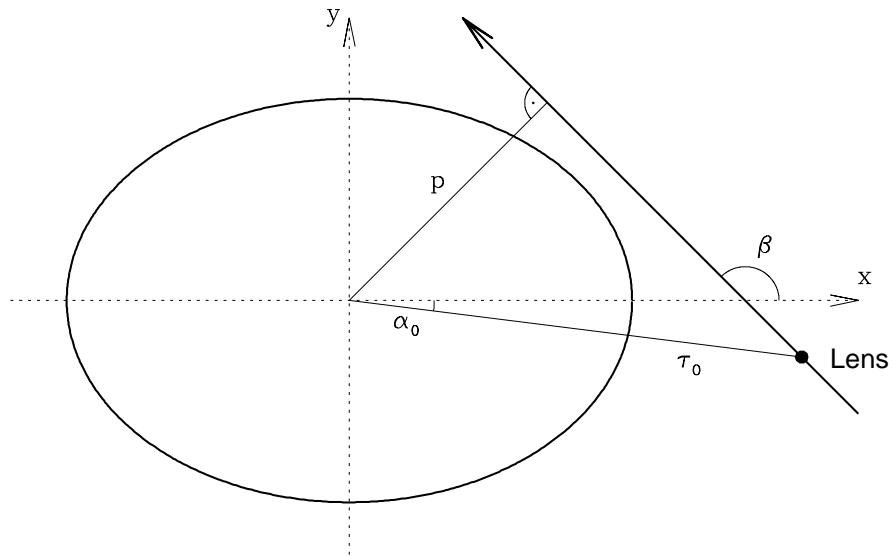}
\caption{Trajectory of a lens in linear motion with respect to the source.}
\end{figure}

\begin{figure}
\plotone{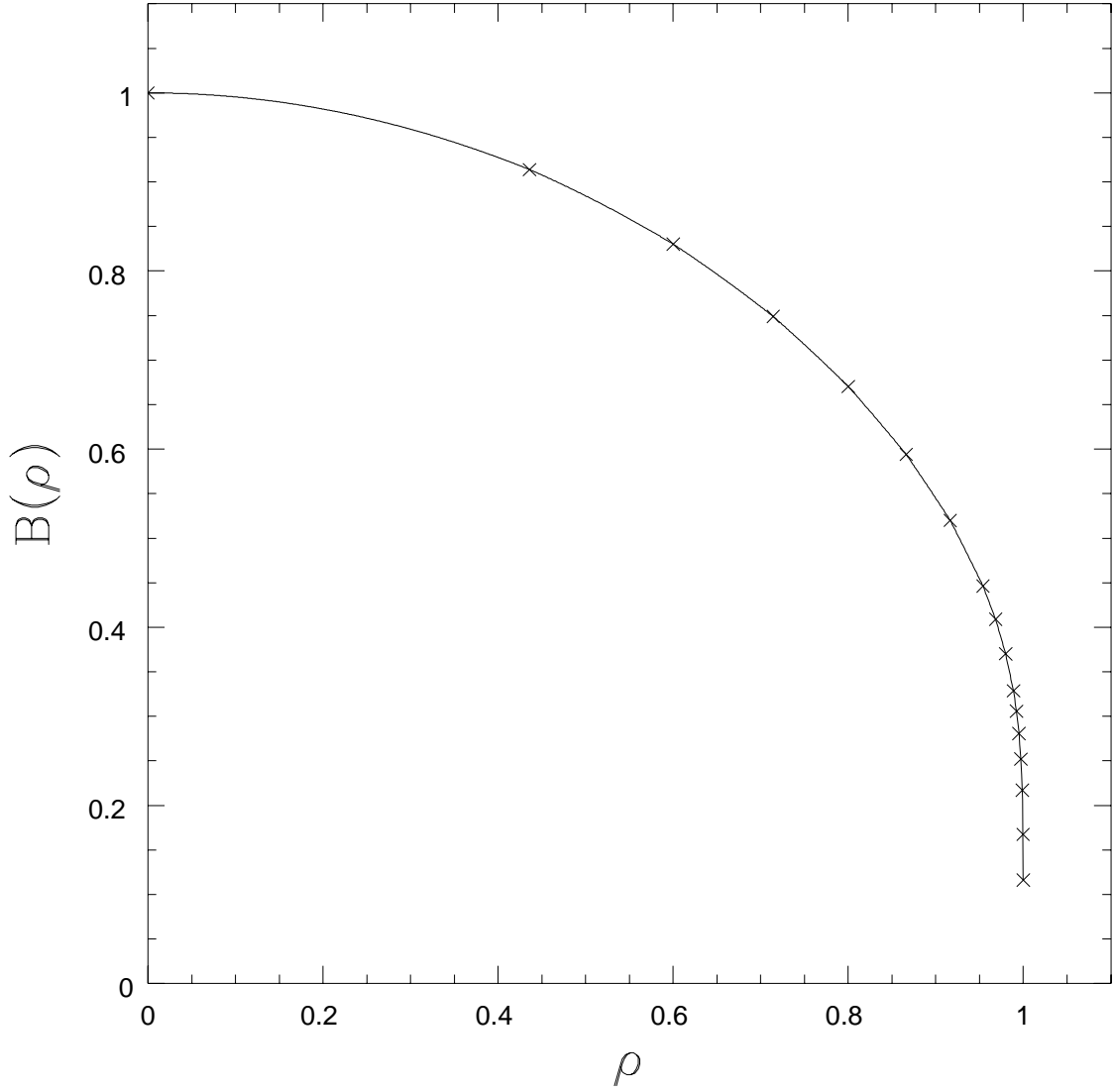}
\caption{Brightness profile of the MACHO 95--30 source in the R-band,
normalized by its central value. The profile is based on a model
atmosphere calculated by Sasselov (1996).}
\end{figure}

\begin{figure}
\plotone{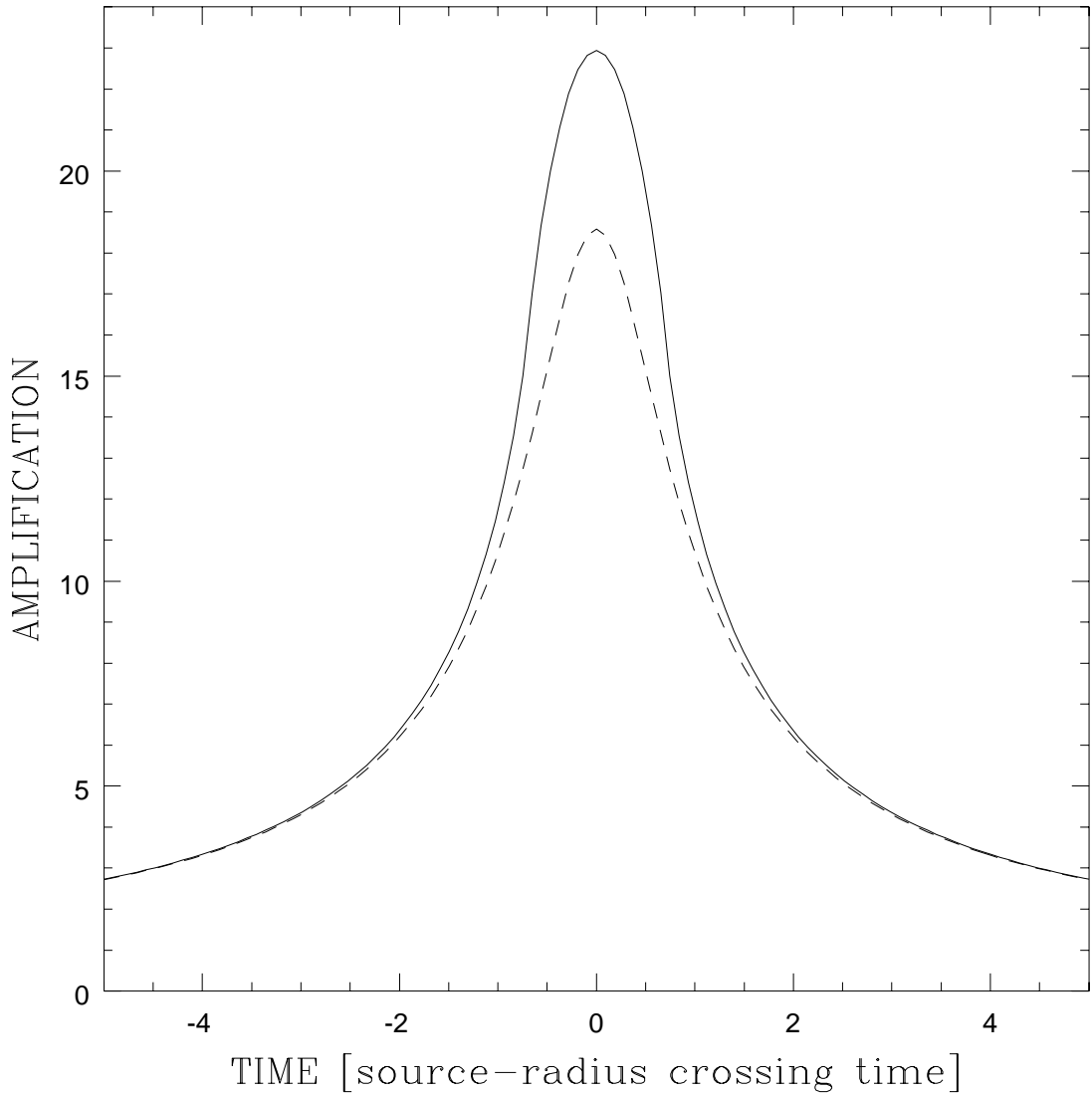}
\caption{Simulated lightcurve of the MACHO 95--30 event (solid line), 
using the brightness profile in Figure 4. 
The lens was chosen to have an Einstein 
radius $\epsilon\approx 13$ and an impact parameter $p\approx 0.7$.
For comparison, the dashed curve 
shows the point--source lightcurve for the same event parameters.
} 
\end{figure}

\begin{figure}
\plotone{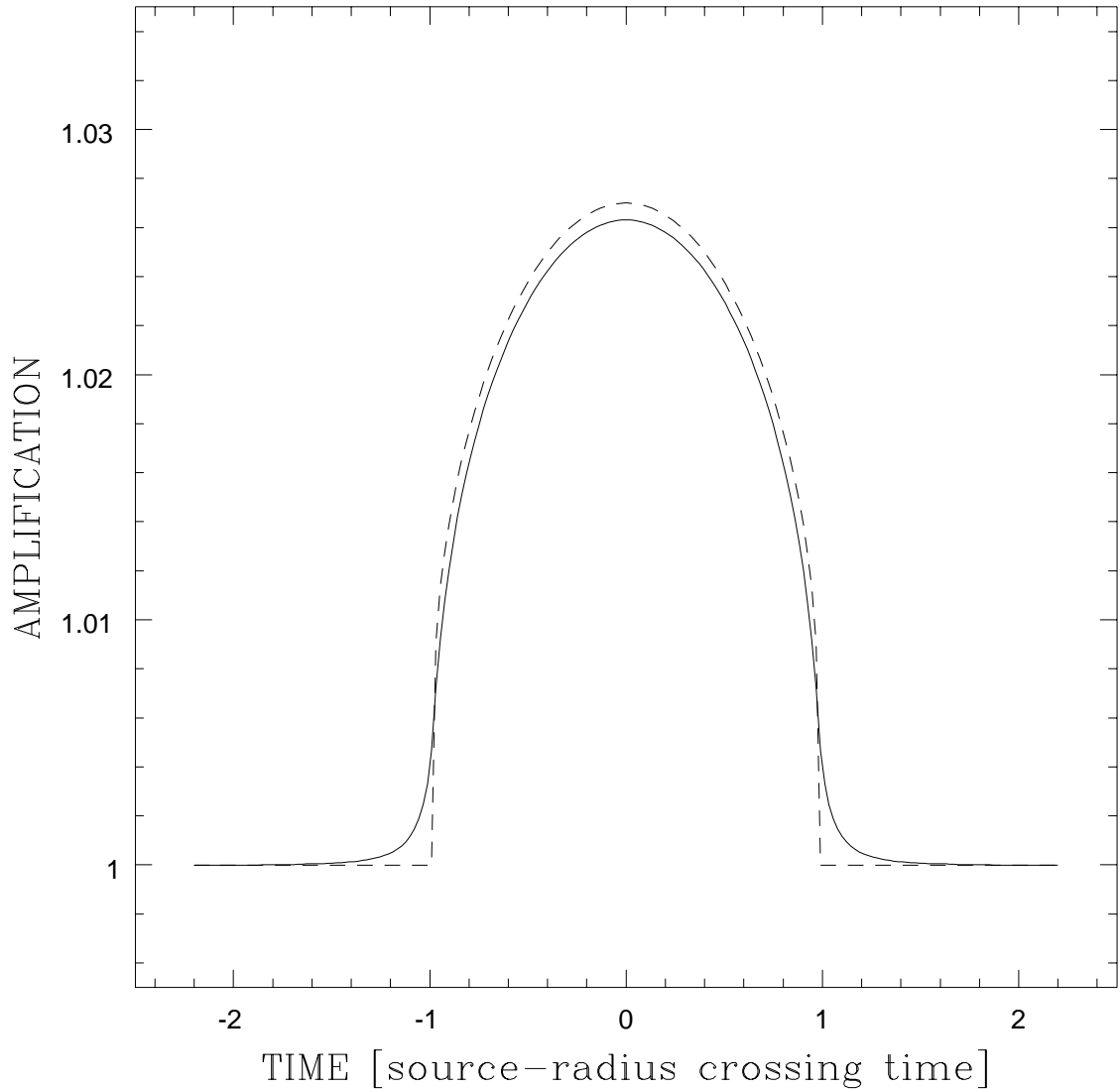}
\caption{Lightcurve of the MACHO 95--30
source for a lens with an Einstein radius $\epsilon=0.1$ and an impact
parameter $p=0.2$.  The exact lightcurve (solid curve) is compared to its
leading order approximation in the low lensing--mass limit (dashed curve).
The brightness profile of the source was taken from Figure 4.}
\end{figure}

\begin{figure}
\plotone{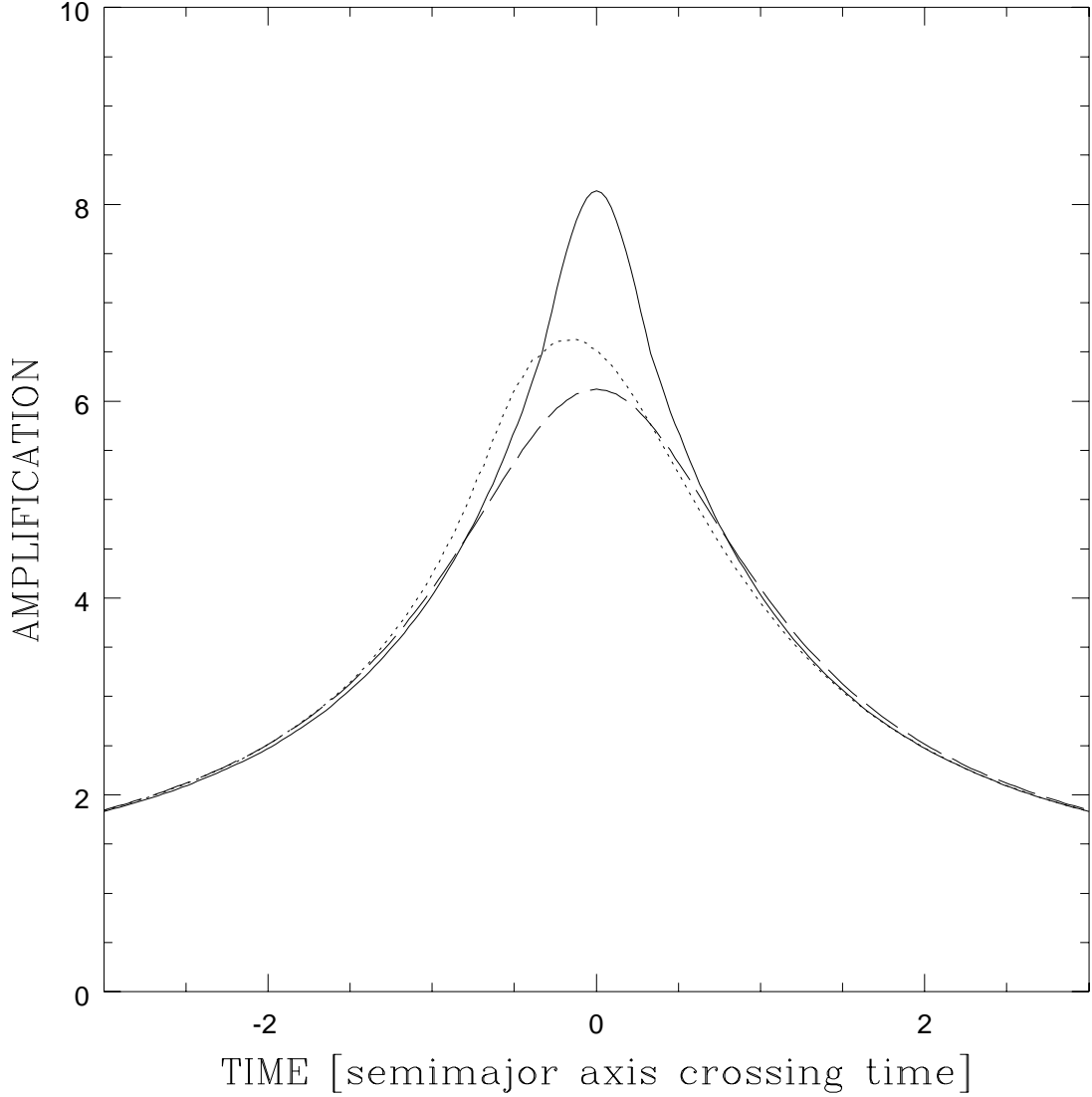}
\caption{Lightcurves for an accretion disk inclined by an angle 
$i=60^\circ$, which is microlensed by a lens with an Einstein radius
$\epsilon=5$. The three curves correspond to events with the same
impact parameter $p=0.8$, but with a different lens orientation.  The solid
line corresponds to a trajectory parallel to the minor axis of the source
($\beta=90^\circ$), the dotted line to a trajectory diagonal to the axes
($\beta=135^\circ$), and the dashed line to a trajectory parallel to the
major axis ($\beta=180^\circ$).  Zero time corresponds to closest approach
in all cases.}
\end{figure}  

\begin{figure} 
\plotone{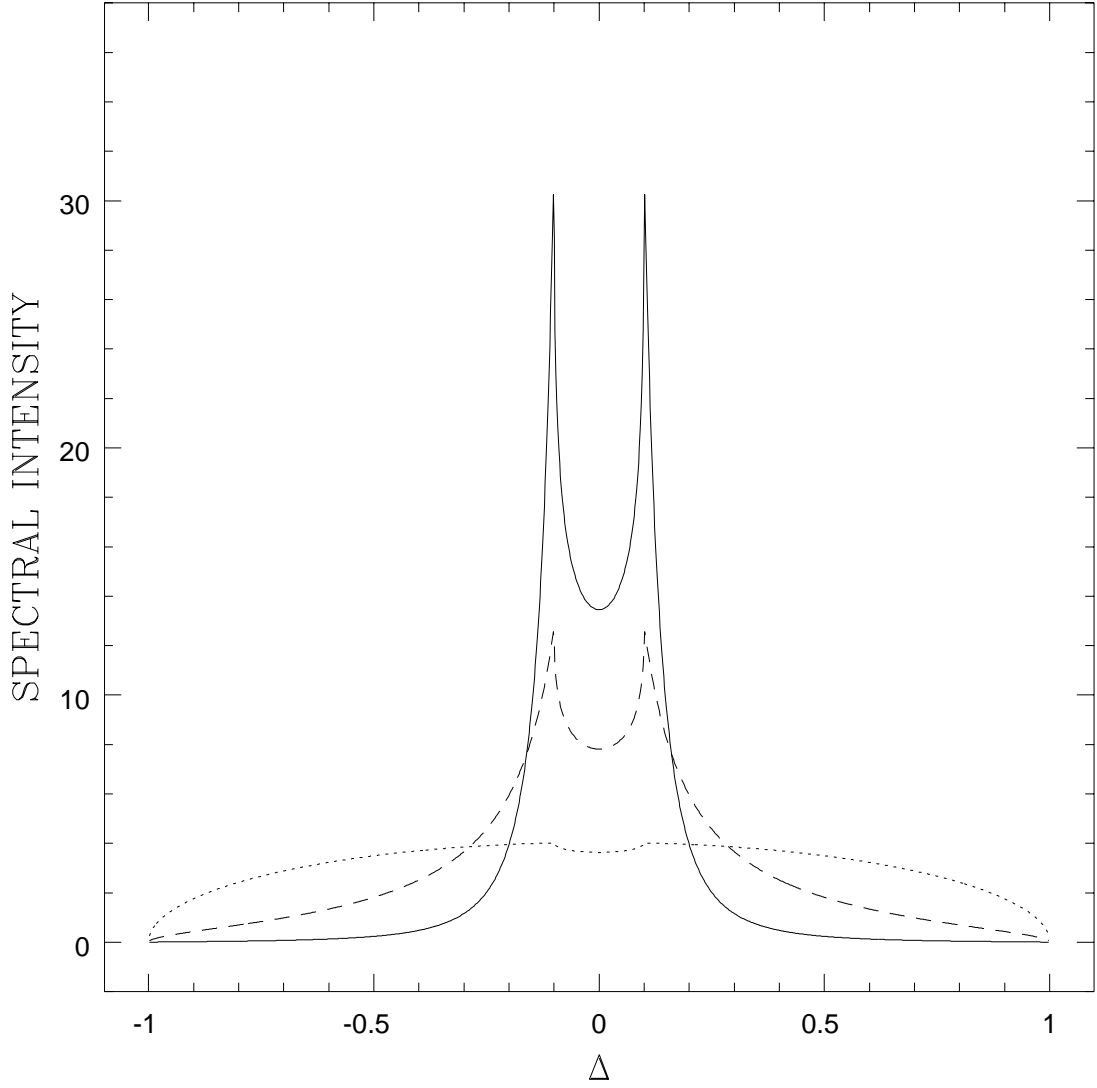} 
\caption{Dependence of the {\it unlensed} profile of a spectral
line emitted by a Keplerian disk, on the radial power--law index of the
disk emissivity. The different line profiles correspond to power--law
indices of $n=-1$ (solid line), $n=-2$ (dashed line) and $n=-3$ (dotted
line), for a disk with an inner radius $\rho_{in}=0.01$ and inclination 
$i=60^\circ$. The vertical axis is in arbitrary units.}
\end{figure}

\begin{figure}
\plotone{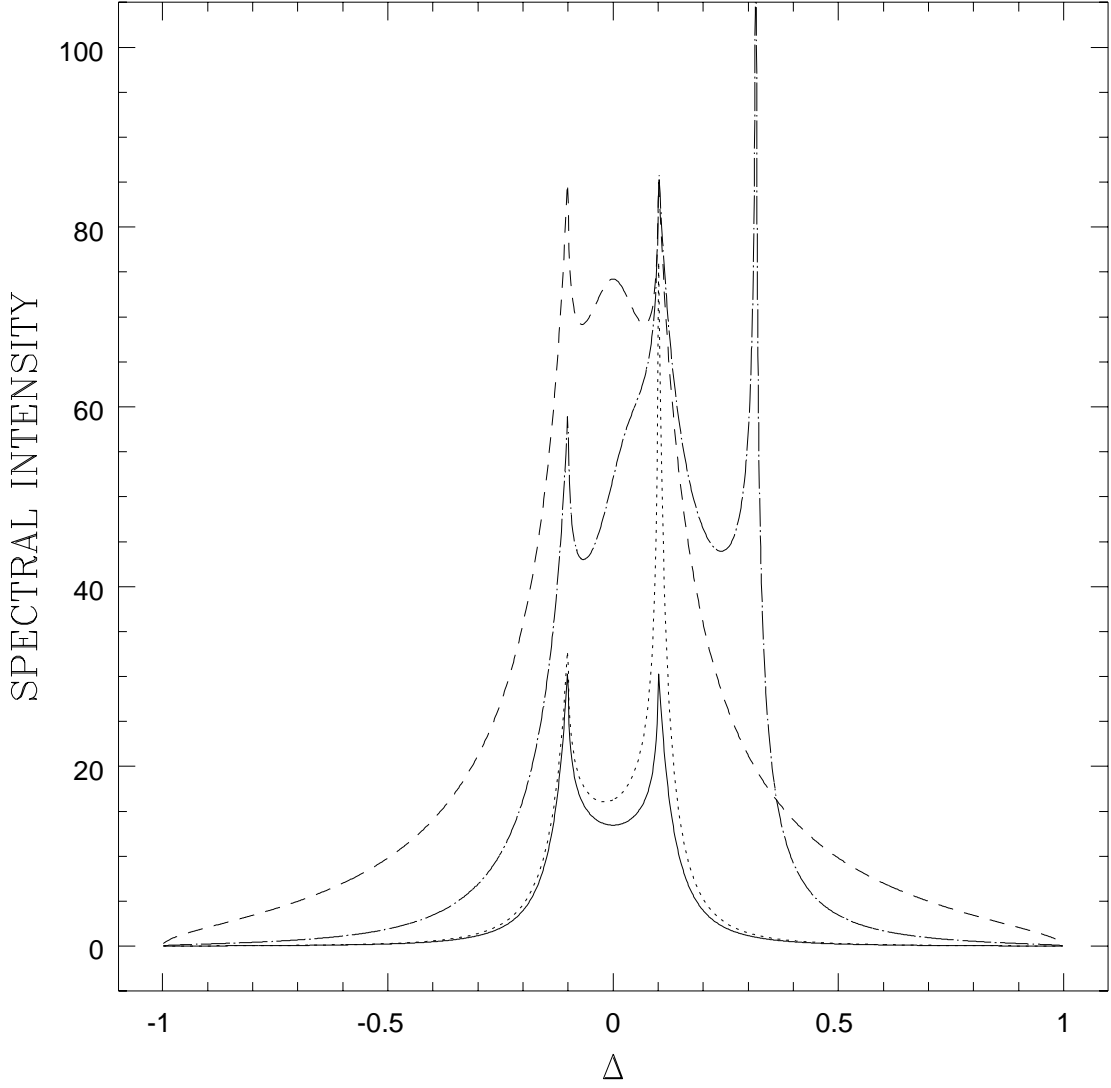}
\caption{Effect of microlensing on the line profile for a disk 
with $n=-1$ in Figure 8, and for various lens positions along the major
axis of the projected disk.  The lens has an Einstein radius
$\epsilon=1$, and is located at $\tau_0=0$ (dashed line), $\tau_0=0.1$
(dot-dashed), and $\tau_0=1.2$ (dotted). The unlensed profile 
is plotted for comparison (solid line).}
\end{figure}

\begin{figure}
\plotone{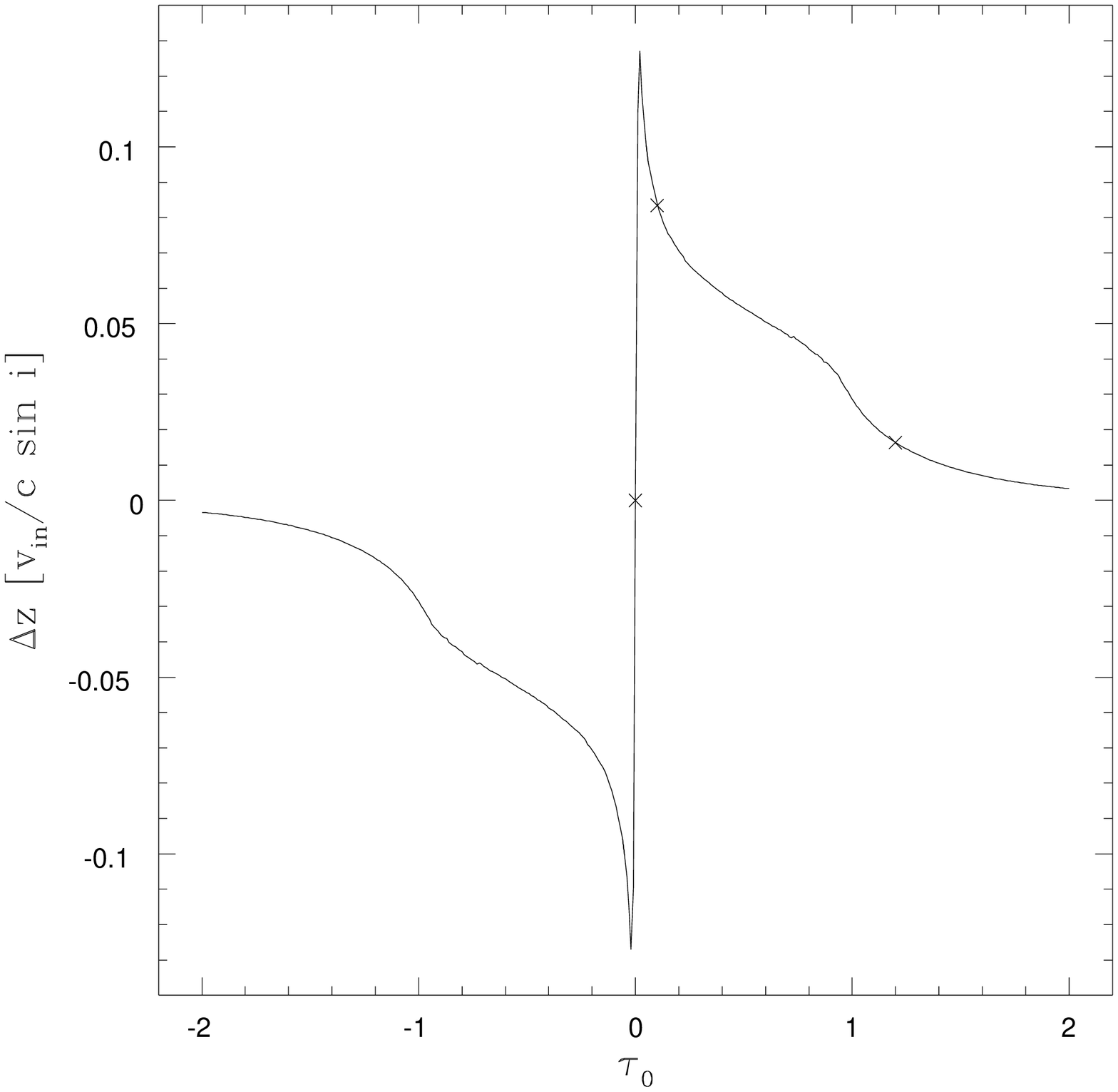}
\caption{Change in mean redshift of the emitted line as a function of lens 
position $\tau_0$ along the major axis of the projected disk. The 
change in redshift is plotted in units of maximum Doppler shift from the 
inner edge of the unlensed disk. Disk parameters are the same as in 
Figures 8 and 9. The three marked points correspond to the lens positions 
for the line profiles in Figure 9.} 
\end{figure}


\begin{references}
\reference{}
Alard, C. 1996, in Proc. IAU Symp. 173, Astrophysical
Applications of Gravitational Lensing, p. 214, Eds. C. S. Kochanek
\& J. N. Hewitt, (Dordrecht: Kluwer Academic Publishers)
\reference{}
Albrow, M., et al. 1996, in Proc. IAU Symp. 173, Astrophysical
Applications of Gravitational Lensing, p. 227, Eds. C. S. Kochanek
\& J. N. Hewitt, (Dordrecht: Kluwer Academic Publishers)
\reference{}
Alcock, C., et al. 1995, ApJ, submitted, preprint  astro-ph/9512146
\reference{}
Alcock, C., et al. 1996, ApJ, submitted, preprint  astro-ph/9606165
\reference{}
Alcock, C., et al. 1997, ApJ, submitted, preprint  astro-ph/9702199
\reference{}
Ansari, R., et al. 1996, A\&A, 314, 94
\reference{}
Bennett, D., et al. 1996, preprint astro-ph/9612208 
\reference{}
Bennett, D., \& Rhie, S. H. 1996, ApJ, 472, 660
\reference{}
Dalcanton, J. J., Canizares, C. R., Granados, A., Steidel, C. C., \& Stocke, 
J. T. 1994, ApJ, 424, 550
\reference{}
Fabian, A. C., Nandra, K., Reynolds, C. S., Brandt,
W. N., Otani, C., Tanaka, Y., Inoue, H., \& Iwasawa, K. 1995, MNRAS, 277, L11
\reference{}
Frank, J., King, A., \& Raine, D. 1985, Accretion Power in Astrophysics 
(Cambridge: Cambridge Univ. Press)
\reference{}
Gould, A. 1995a, ApJ, 447, 491
\reference{}
Gould, A. 1995b, ApJ, 455, 37
\reference{}
Gould, A., \& Miralda-Escud\'e, J. 1996, preprint astro-ph/9612144
\reference{}
Gould, A., \& Welch, D. L. 1996, ApJ, 464, 212
\reference{}
Heyrovsk\'y, D., Loeb, A., \& Sasselov, D. 1996, in Proc. IAP Colloquium
12, Variable Stars and the Astrophysical Returns of Microlensing Surveys,
eds. R. Ferlet and J. P. Maillard, (Paris: Editions Frontieres)
\reference{}
Irwin, M. J., Webster, R. L., Hewett, P. C., Corrigan, R. T., \&
Jedrzejewski, R. I. 1989, AJ, 98, 1989
\reference{}
Iwasawa, K., et al. 1996, MNRAS, 282, 1038
\reference{}
Jaroszy\'nski, M., Wambsganss, J., \& Paczy\'nski, B. 1992, ApJ, 396, L65
\reference{}
Laor, A. 1990, MNRAS, 246, 369
\reference{}
Lennon, D. J., Mao, S., Fuhrmann, K.,
\& Gehren, T. 1996, ApJ, 471, L23 
\reference{}
Loeb, A., \& Sasselov, D. 1995, ApJ, 449, L33
\reference{}
Mao, S., et al. 1994, Bull. American Astron. Soc., 185, \#17.05
\reference{}
Maoz, D. 1996, to appear in the proceedings of IAU Colloquium 159,
Shanghai, June 1996, preprint astro-ph/9609174
\reference{}
Maoz, D., \& Gould, A. 1994, ApJ, 425, L67
\reference{}
Miyoshi, M., Moran, J., Herrnstein, J., Greenhill, L., Nakai, N., Diamond,
P., \& Inoue, M. 1995, Nature, 373, 127
\reference{}
Nemiroff, R. 1988, ApJ, 335, 593
\reference{}
Paczy\'nski, B. 1986, ApJ, 304, 1
\reference{}
Perna, R., \& Loeb, A. 1997, ApJ, submitted, astro-ph/9701226
\reference{}
Peterson, B. M. 1993, PASP, 105, 247
\reference{}
Pratt, M. R. 1996, in collection of talks in the 2nd International 
Workshop on Gravitational Microlensing Surveys, LAL-Orsay, p. 243
\reference{}
Pratt, M. R., et al. 1996, in Proc. IAU Symp. 173, Astrophysical
Applications of Gravitational Lensing, p. 221, Eds. C. S. Kochanek
\& J. N. Hewitt, (Dordrecht: Kluwer Academic Publishers)
\reference{}
Racine, R. 1991, AJ, 102, 454
\reference{}
Rauch, K. P., \& Blandford, R. D. 1991, ApJ, 381, L39
\reference{}
Sasselov, D. 1996, in Proc. IAP Colloquium 12, Variable Stars and the
Astrophysical Returns of Microlensing Surveys, eds. R. Ferlet and
J. P. Maillard, (Paris: Editions Frontieres)
\reference{}
Sasselov, D., Heyrovsk\'y, D., \& Loeb, A. 1997, in preparation
\reference{}
Schild, R. E., \& Smith, R. C. 1991, AJ, 101, 813
\reference{}
Schneider, P., Ehlers, J., \& Falco, E. E. 1992, 
Gravitational Lenses (Berlin: Springer-Verlag)
\reference{}
Schneider, P., \& Wambsganss, J. 1990, A\&A, 237, 42
\reference{}
Steidel, C. C., Pettini, M., Dickinson, M., \& Persson, S. E.
1994, AJ 108, 2046
\reference{}
Steidel, C. C., Bowen, D. V., Blades, J. C., \& Dickinson, M.
1995, ApJ, 440, L45
\reference{}
Steidel, C. C., Giavalisco, M., Dickinson, M., \& Adelberger, K. L. 1996, 
AJ, 112, 352 
\reference{}
Tanaka, Y., et al. 1995, Nature, 375, 659
\reference{}
Udalski, A., et al. 1994, Acta Astron., 44, 165
\reference{}
Wambsganss, J. 1990, Gravitational Microlensing, report MPA 550, Garching
\reference{}
Wambsganss, J., \& Kundic, T. 1995, ApJ, 450, 19
\reference{}
Wambsganss, J., Paczy\'nski, B., \& Schneider, P. 1990, ApJ, 358, L33
\reference{}
Warner, B. 1995, Cataclysmic Variable Stars, (Cambridge: Cambridge Univ. 
Press)
\reference{}
Witt, H. J. , \& Mao, S. 1994, ApJ, 430, 505

\end{references}
\end{document}